\documentclass[a4paper,12pt]{article}

\usepackage{amsmath}
\usepackage{graphicx}

\begin{document} 

\title{ON STATISTICAL MECHANICS DEVELOPMENTS OF CLAN CONCEPT IN MULTIPARTICLE PRODUCTION}
\author{M. Brambilla (*), A. Giovannini (**), R. Ugoccioni (**)\\
 \textit{(*) Dipartimento di Fisica Nucleare e Teorica and INFN - Sez. di Pavia}\\
 \textit{Via A. Bassi 6, 27100 Pavia, Italy} \\
 \textit{(**) Dipartimento di Fisica Teorica and INFN - Sez. di Torino}\\
 \textit{Via P. Giuria 1, 10125 Torino, Italy}}
\maketitle

\begin{abstract}

Clan concept has been introduced in multiparticle dynamics in order to interpret the wide occurrence of
negative binomial (NB) regularity in $n$-charged particle multiplicity distributions (MDs) in various high energy collisions. \\
The centrality of clan concept led to the attempt to justify its occurrence within a statistical model of clan formation and evolution. In this framework all thermodynamical potentials have been explicitely calculated in terms of NB parameters. Interestingly it was found that NB parameter $k$ corresponds to the one particle canonical partition function. \\
The goal of this paper is to explore a possible temperature $T$ and volume $V$ dependence of parameter $k$ in various classes of events in high energy hadron-hadron collisions. \\
It is shown that the existence of a phase transition at parton level from the ideal clan gas associated to the semihard component with $k>1$ to the ideal clan gas of the hard component with $k<1$ implies a discontinuity in the average number of particles at hadron level.

\end{abstract}

\section{Introduction. A recollection of statistical and quantum theoretical ideas in multiparticle production}

One of the most interesting experimental facts in the thierties of the past century had been the discovery in cosmic ray physics of multiparticle production from a primary hadron of given energy. \\
E.\ Fermi \cite{Fermi} in order to explain the new phenomenon proposed a thermodynamical approach to it. In the same years L.\ Landau \cite{Landau} focused his attention on the hydrodynamical aspects of the production process, whereas W. Heisenberg \cite{Heisenberg} was impressed by its non-linearity. Accordingly, Heisenberg took a different point of view from Fermi and Landau and argued that the natural framework for the description of multiparticle production should be found not in statistical mechanics but in a non-linear field theory of a "new nuclear force". \\
In between these two quite extreme approaches, the first one inspired by statistical mechanics and the second one by quantum field theory, are contained all the later developments of multiparticle production in high energy physics. Today we know indeed that the non-linear quantum field theory foreseen by Heisenberg is Quantum Chromo Dynamics (QCD), the quantum field theory of the strong interactions (the "new nuclear force"...). \\
At the same time the discovery in the sixties of the past century of unsuspected regularities in final particle multiplicity distributions originated by primary hadrons with different initial energies when plotted vs.\ $n$ in cosmic ray physics by P.K. Mac Keown and A.W. Wolfendale \cite{Wolfendale} pointed out the statistical aspects of collective variables properties like normalized cumulants ($K_n$) and factorial moments ($F_n$) of the final particle multiplicity distribution describing the production process. It was found in fact that $n$ final pion multiplicity distributions, $P_n$, when plotted vs.\ $n$ turned out to be quite well fitted by a NB (Pascal) MD with the average multiplicity, $\bar n$, and $k = \frac{{\bar n}^2}{D^2-\bar n}$ ($D$ is the dispersion of the MD) as characteristic parameters. In addition, $\bar n$ is becoming larger and parameter $k$ smaller as the energy of the primary hadron increases. \\
One of the main results along this line of search -- it should be mentioned -- had been in view of the decreasing of parameter $k$, as the primary energy increased, clearly KNO scaling violation. \\
Once again cosmic ray physics findings anticipated future discovery in the accelerator region. \\
It has to be noticed in fact that the study of collective variables properties in various classes of high energy collisions at various c.m.\ energies and in different (pseudo)rapidity intervals in the accelerator region performed in the seventies and in the eighties by various Collaborations revealed in $pp$, $p \bar p$, $\mu p$ and amazingly in $e^+ e^-$ annihilation the same general trends and the same regularities discovered in cosmic ray physics. \\
In order to reproduce these experimental facts at final hadron level starting from perturbative QCD at parton level it is well known that one has to rely in a region where QCD has no predictions on Monte Carlo event generators based on DGLAP evolution equations with an ad hoc hadronization prescription. This last line of search has to be considered not fully satisfactory in view of the quite large number of parameters involved in the event generator models and suggests to dig more carefully in the statistical mechanics aspects of the approximate universal regularities discovered in $P_n$ vs.\ $n$ plots with a limited number of well identified physical parameters. The goal in this framework is to understand within the universality of the NB (Pascal) regularity the statistical counterpart of the universality of QCD in all explored high energy collisions. \\
Accordingly, it is instructive to start with a critical reading of the main statistical mechanics literature on multiparticle production in addition to the pioneering work of Fermi and Landau. \\
Among them Feynman fluid analogy should be mentioned. As discussed in \cite{thermo-clan} here the cross-section for the production of $n$ final particles plays the role of the partition function in the canonical ensemble and it is an integral over phase space of the square of a matrix element, which corresponds to the Gibbs distribution $e^{-H/k_{B}T}$ with Hamiltonian $H$, Boltzmann factor $k_B$ and temperature $T$; volume $V$ is identified by Feynman with the extension of phase space and fugacity $z$ with the dummy variable $u$ appearing in the definition of the $n$-particle multiplicity $p(n)$ generating function $G(u)$
$$
G(u) \equiv \sum_{n}u^n p(n).
$$
It is to be remarked that this idetification is again rather unsatisfactory from our point of view because one has to satisfy at the same time the definitions of the aveage number of particles from the grand canonical ensemble and from the definition of the generating function. These definitions can be satisfied at the same time only in the limit of zero chemical potential. \\
Another approach was proposed by Scalapino and Sugar: they defined the probability amplitude to produce a particle at rapidity $y$, denoted by $\Pi(y)$, as a random field variable, then they introduced a functional $F[\Pi]$ which played a role analogous to the free energy for a system in thermal equilibrium. One can then obtain the $n$-particle inclusive distribution by averaging the product of the squares of the amplitude, $\Pi^2(y_1) \cdots \Pi^2(y_n)$, with a weight given by $e^{-F[\Pi]}$.
Lacking the knowledge necessary to calculate $F[\Pi]$ from the underlying dynamics, the authors parametrized it in retaining the first three terms in a series expansion, then solved the problem in few particular cases. Remarkably, to leading order in the size of the allowed full rapidity range, they obtain a generating function of an infinite divisible distribution (IDD), a class of distributions to whom the NB (Pascal) MD is belonging. \\
More recent results concerning KNO scaling violation and phase transitions can be founf in the literature (see Ref.s quoted in \cite{thermo-clan}). \\
Stimulated by these findings some of the present Authors in collaboration with S. Lupia proposed a simplified approach to multiparticle production in high energy collisions based on the NB (Pascal) MD and valid for any chemical potential. The interpretation of the NB regularity was given in terms of clan structure analysis (see Ref.\cite{thermo-clan} and Section 2 of the present paper). \\
The plan of the paper is the following. \\
After introductory remarks discussed in Section 1, in Section 2 the interpretation of the NB (Pascal) MD regularity in multiparticle production in terms of clan structure analysis and its more recent applications to the components (or classes of events) in high energy collisions are reviewed. \\
In Section 3 a summary of previous results on clan thermodynamics is given. \\
In Section 4 the explicit dependence on temperature $T$ and volume $V$ -- not shown in \cite{thermo-clan} -- of the one particle canonical partition function $k$, i.e.\ of the NB parameter $k$ in the various components (soft, semihard and hard) of $pp$ collisions, is determined. \\
In Section 5 it is shown that the existence of a phase transition at parton level from the ideal clan gas associated in $pp$ to the semihard component with $k > 1$ to the ideal clan gas associated to the hard component with $k < 1$ implies a discontinuity in the average number of particles of the two mentioned components at final hadron level. \\
In the Appendix a theorem on the zeroes of a degree $M$ polynomial used in Section 5 is proved.

\section{Clan concept and Negative Binomial (NB) regularity in high energy multiparticle production}

In order to interpret the wide occurrence of negative binomial $n$ charged particle multiplicity distribution, $P_{n}^{[NB]}$, regularity in various classes of high
energy collisions in the eighties it has been proposed an alternative parametrization to standard NB parameters, i.e.\ the average charged multiplicity $\bar n$ and parameter $k$ of $P_{n}^{[NB]}=\frac{k (k+1) ... (k+n-1)}{n!}{\left(\frac{\bar n}{\bar n + k}\right)}^n{\left(\frac{k}{\bar n + k}\right)}^k$. The proposed new parameters were the average number of groups of particles of common ancestor (called in high energy multiparticle phenomenology average number of clans, $\bar N$), and the average number of particles per clan, 
$\bar n_c$. It has been also shown that the new parameters are linked to the standard ones by the following not trivial relations
\begin{equation}
\bar N = k \ln\left(1+\frac{\bar n}{k}\right)  \mbox{ and } \bar n_c=\bar n/\bar N.
\end{equation}
Notice that $\bar n_c$ is a scaling function of $\bar n/k$. \\
It turns out that clans in various classes of events in a collision are independently produced whereas particles within each clan are distributed according to a logarithmic multiplicity distribution and all multiplicity correlations are exhausted within each individual clan. These properties can be expressed by using the corresponding generating functions of the NBMD in the two mentioned sets of parameters $(\bar n,k)$ and $(\bar N,\bar n_c)$. 
In fact
\begin{equation}
G^{[NB]}(\bar n,k;x)=\sum_{n=0}^{\infty}P_{n}^{[NB]}\,x^n=\left[\frac{k}{k+\bar{n}(1-x)}\right]^k
\end{equation}
and
\begin{equation}
G^{[NB]}(\bar N,\bar n_c;x)=\exp [\bar N (G_{ln}(\bar n_c;x)-1)] 
\end{equation}
with $G_{ln}(\bar n_c;x)=\frac{\ln(1-zx)}{\ln(1-z)}$ and $z=\frac{\bar n}{\bar n + k}$. \\
In addition it should be remembered that $1/k$ is the ratio of the probability to have two particles within the same clan over the probability to have in a  statistical framework the two particles in two distinct clans, i.e.\ $1/k$ is an aggregation parameter which is linked to the two-particle correlation function in rapidity variables $y_1$ and $y_2$, $C_2(y_1,y_2)$, by $\frac{1}{k}=\int C_2(y_1,y_2)\,dy_1 dy_2$. \\
The introduction of clan concept within a single NBMD allowed to single out the main differences in multiparticle production in $e^+ e^-$ annihilation, deep inelastic scattering and hadron-hadron collisions.
Experts started to talk of NB universality in multiparticle production. The regularity at final hadron level was supported also by QCD in view of the
fact that NBMD occurs at the parton level in the leading log approximation with a fixed cut-off regularization prescription in order to describe quark- and
gluon-jets as Markov branching processes. Notice that at parton level $1/k$ is the ratio of the $g \rightarrow g + g$ vertex over the $q \rightarrow  q + g$ vertex, and in the $(\bar N,\bar n_c)$ NB parametrization $\bar N$ corresponds to the average number of bremsstrahlung gluon-jets, as shown in Ref.\cite{Giovannini}. \\
After initial enthusiasm which led to a huge list of sound phenomenological papers in multiparticle production in high energy collisions, it was found by UA5 collaboration at CERN $p \bar p$ collider that the regularity was violated at top c.m.\ energy (900 GeV) and by DELPHI collaboration in $e^+ e^-$ annihilation at LEP c.m.\ energy: a shoulder structure occurred in the total $n$ charged particle multiplicity distribution, $P_n$, when plotted versus $n$, which a single $P_{n}^{[NB]}$ was unable to reproduce. Interestingly the regularity violated in the total sample of events in both reactions was restored at a deeper level of investigation, i.e.\ in terms of its components (or substructures or classes of events) general properties: it was found indeed that the regularity initially discovered in the low c.m.\ energy domain and then violated in the total sample of events at higher c.m.\ energy, continued to be valid in the new energy region when applied to the various $i$-cl!
 asses of events ($i=1,2,3...$) with characteristic $\bar n_i$, $k_i$ or $\bar N_i$, $\bar n_{c,i}$ parameters. Furthermore, it was found that the weighted superposition of the various $i$-classes of events observed in $pp$ collisions and in $e^+ e^-$ annihilation, whose $n_i$ particles multiplicity distribution was described by a $P_{n_i}^{[NB]}$ with characteristic and in general different NB parameters for each $i$-class of events, allowed to interpret not only shoulder structure in the total sample of events when plotted versus $n$, but also experimentally observed $H_q$ oscillations when plotted versus $q$ ($H_q$ is the ratio of $q$-factorial cumulants, $K_q$, and $q$-factorial moments, $F_q$, of the corresponding total multiplicity distribution) as well as forward-backward $n$-particle multiplicity correlation strenght behaviour in the total sample of events. \\
All these facts stimulated the attempt to justify the introduction of clan concept and related NB regularity in the various classes of events occurring in different high energy collisions in a thermodynamical framework. Clan thermodynamics was then proposed \cite{thermo-clan}.

\section{A summary of previous results on clan thermodynamics}

In paper \cite{thermo-clan} starting from the relationship between the grand canonical partition function $\mathcal{Q}(z,T,V)$ of a system with fugacity $z$, volume $V$, temperature $T$ and the $n$-particle (parton) canonical one, $Z_n$, for the Negative Binomial Multiplicity Distribution (NBMD) with the average multiplicity $\bar n$ and $k$ as characteristic NB parameters, it was found that
\begin{equation}
\mathcal{Q}(z,T,V)=\sum_{n=0}^{\infty}z^n Z_n(T,V)={\left(\frac{k}{\bar n + k}\right)}^{-k}=e^{\bar N}
\end{equation}
where
\begin{equation}
z=e^{\frac{\mu}{k_B T}}=\frac{\bar n}{\bar n + k}
\end{equation}
with $\mu$ the chemical potential, $Z_n = \frac{k (k+1) ... (k+n-1)}{n!}$ and the probability of finding $n$ particles (partons) $p(n)$ in the system is given by
\begin{equation}
p(n)=\frac{z^n Z_n(T,V)}{\sum_{n=0}^{\infty}z^n Z_n(T,V)}=z^n Z_n(T,V) \, p(0)
\end{equation}
with $p(0)={\left(\frac{k}{\bar n + k}\right)}^k$ and $Z_0 = 1$. Notice that particle (parton) fugacity $z$ satisfies the condition $0 \leq z \leq 1$, i.e.\ our particle (parton) system behaves as a real boson gas. It should be pointed out that in our approach, as in those mentioned in the Introduction, we are always working in the framework of equilibrium thermodynamics. \\
It turns out that the canonical partition function for a system of one particle $Z_1$ corresponds exactly to parameter $k$, i.e.\ to an unknown function of temperature $T$ and volume $V$
\begin{equation*}
k \equiv k(T,V).
\end{equation*}
Since the thermodynamical potentials calculated in Appendix A of Ref.\cite{thermo-clan} have been shown to depend on the partial derivative of $k$ with respect to temperature $T$, the main goal of the present paper is the determination of the explicit dependence of $k$ on temperature $T$ and volume $V$. \\
This search is interesting also in view of the fact discussed in Ref.\cite{complex} that NB parameter $k$ controls the geometry of the zeroes distribution in the complex $z$-fugacity plane of the $M$-order algebraic equation of the $M$-truncated grand canonical partition function
\begin{equation}
\mathcal{Q}_M(z,T,V)=\sum_{n=0}^{M}z^n Z_n(T,V)=0.
\end{equation}
In fact, it has been shown in Ref.\cite{complex} that in the rescaled variable $u=zx$ for $k$ larger than one all zeroes stay in the complex rescaled $u$-plane inside the circle of unit radius $|u| = 1$, whereas for $k$ smaller than one the zeroes stay outside the circle $|u| = 1$ and on the circle itself for $k=1$. \\
Accordingly,
\begin{equation}
\ln\mathcal{Q}(z,T,V) = \bar N 
\end{equation}
or in terms of the Boltzmann equation of state
\begin{equation}
P\,V = k_{B}T \ln\mathcal{Q} = k_{B}T\bar{N} 
\end{equation}
with $P$ the pressure of the ideal gas of clans and $k_{B}$ the Boltzmann constant. Volume $V$ is here the space volume occupied by $\bar N$ clans of the ideal clan gas satisfying Boltzmann equation of state (9). \\
Notice that here we consider clans and not particles (or partons) as in Eq.(5). Accordingly, our system can be interpreted as a \emph{real boson gas} with fugacity $z$ given by Eq.(5) when we are dealing with particles (partons) or as an \emph{ideal Boltzmann gas} when we are dealing with clans. \\
These findings have an intriguing application to the eventual new class of hard events foreseen at LHC when $\bar N \rightarrow 1$ \cite{hard}. In this class of events $k$ is less than one, whereas it is larger than one for the soft and semihard classes of events. The point to be cleared up concerns indeed the possibility that the clan gas in going from $k$ larger than one (i.e.\ from the semihard clan gas) to $k$ smaller than one (i.e.\ to the hard clan gas) undergoes a phase transition and the nature of the phase transition itself. \\
Notice that results in the hadronic ($h$) sector of the collision can be extended to the partonic ($p$) one via the generalized local parton($p$)-hadron($h$) duality (GLPHD), i.e.\ by assuming on phenomenological grounds that $k_p = k_h$ and $\bar n_h =\sigma \bar n_p$ with $\sigma \simeq 2$ (see Ref.\cite{libro} and related bibliography). \\
The existence of the first two classes of events in $e^+ e^-$ annihilation (class I: $2$-jet events; class II: $3$- or more-jet events) as well as in $pp$ collisions (class I: events without minijets; class II: events with minijets) was motivated by the discovery in both reactions of three relevant facts extensively discussed in Ref.s \cite{clan-def} and \cite{forward-backward} and mentioned in the introduction of the present paper: \\
a. the shoulder structure in $n$-particle multiplicity distribution of the total sample of events $P_n$ when plotted vs.\ $n$; \\
b. $H_q$ vs.\ $q$ oscillations, with $H_q$ the ratio of $q$-factorial cumulants, $K_q$, and the $q$-factorial moments, $F_q$, of the corresponding MD; \\
c. the lack of forward (F) to backward (B) $n$-particle multiplicity correlation in $e^+ e^-$ annihilation and its strong enhancement in $pp$ collisions. \\
It is to be pointed out that all these collective variable properties were explained in terms of the weighted superposition of different classes of events, each described by a NB (Pascal) MD with characteristic NB parameters. \\
The explanation of point c. is very probably a consequence of a huge colour exchange mechanism occurring at parton level in $pp$ collisions and of a very weak colour exchange in $e^+ e^-$ annihilation. Clan structure analysis in terms of the new parameters $\bar N$ and $\bar n_c$ revealed in this framework all its remarkable implications in the dynamics of a collision: clans turned out to be more numerous in $e^+ e^-$ annihilation but less populated, whereas just the opposite occurred in $pp$ collisions; here clans are less numerous than in $e^+ e^-$ annihilation, but more populated. An intermediate situation was shown to happen in deep inelastic scattering, i.e.\ the average number of clans is less numerous than in $e^+ e^-$ annihilation but the average number of particles (partons) per clan larger. \\
In conclusion, the dynamics of the collision in terms of $\bar N$ and $\bar n_c$ was understood as a two step process: clans were firstly independently produced (each clan contains at least one particle or parton, its ancestor), each clan then decay accordingly to parton or hadron cascading mechanism with all particles (partons) correlations exhausted within each clan. \\
All these facts led to the conjecture that introduced clan concept is not only the way to a new useful and pictorial parametrization to be added to the standard one in terms of $k$ and $\bar n$, but it points in the direction of clans as true physical objects to be looked for in future experimental work. In addition, the reduction with the increase of the c.m.\ energy of $\bar N_{semihard}$ observed in $pp$ collisions \cite{semihard} arises interesting questions on the properties of the eventual third class of quite hard events, characterized by $\bar N_{hard} \rightarrow 1$, $k_{hard}\ll1$ and $\bar n_{hard}$ very large with respect to $\bar n_{soft}$ and $\bar n_{semihard}$. \\
By assuming that the third class of events is blowing up already at LHC c.m.\ energy (14 TeV), we distinguish at such energy in the collision at least three classes of events whose weighted superposition is describing collective variable behaviours (i.e.\ of $P_n$, $K_n$, $F_n$, FB multiplicity correlations) in the full sample of events. \\
The three expected classes of events in $pp$ are therefore (see Ref.\cite{Walker}): \\
I.  a soft one (events with no minijets) due to one parton-one parton scattering; \\
II. a semihard one (events with minijets) due to two partons-two partons scattering; \\
III. a very hard one originated by three partons-three partons scattering, with $\bar N_{hard} \rightarrow 1$, $k_{hard}\ll1$ and
$\bar n_{hard} \gg \bar n_{semihard} > \bar n_{soft}$. \\
Notice that according to Ref.\cite{Walker} temperatures as well as the parton densities of the $i$-classes are expected to increase with the hardness of the collision. The reduction of the average number of clans as the temperature increases (and the parton collisions become more numerous) suggests indeed the starting of a higher parton density process. \\
In the present paper we want to explore the onset of the hard component at parton level from the semihard one in $pp$, assuming it occurs at LHC in the framework of clan thermodynamics discussed in \cite{thermo-clan}, and in particular to determine the $T$ and $V$ dependence of NB parameter $k$.

\section{Temperature and volume dependence of the one particle canonical partition function $Z_1=~k$}

In order to calculate analitically the temperature dependence of NB parameter $k$, we decided to use consistency relation between the entropy expressed in terms of the grand canonical partition function $\mathcal{Q}$ 
\begin{equation}
S=k_{B}\ln\mathcal{Q}+k_{B}T \left(\frac{\partial}{\partial T} \ln\mathcal{Q} \right)_{\mu,V},    
\end{equation}
($\mu$ is here the chemical potential, which is linked to fugacity $z$ as already pointed out by $z=e^{\frac{\mu}{k_B T}})$ and the entropy given in the Appendix of Ref.\cite{thermo-clan}
\begin{equation}
S = \frac{U-F}{T} = k_{B}\bar{N}-\bar{n}\frac{\mu}{T}+k_{B}\frac{\bar{N}}{k} T \left(\frac{\partial k}{\partial T} \right)_{V},    
\end{equation}
where
\begin{equation*}
U = k_{B}T^2 \left(\frac{\partial \bar N}{\partial T}\right)_{z,V} = k_{B}T^2 \frac{\bar N}{k}\left(\frac{\partial k}{\partial T}\right)_V,    
\end{equation*}
\begin{equation*}
F = \bar n \,k_{B}T \ln \left(\frac{\bar n}{\bar n + k}\right) - k_{B}T \,k \ln \left(1 + \frac{\bar n}{k}\right),    
\end{equation*}
and
\begin{equation*}
\bar N = \ln\mathcal{Q} = -k \ln(1-z).    
\end{equation*}

It should be noticed that partial derivative with respect to the temperature $T$ at constant $\mu$ and $V$ can be written in terms of fugacity $z$ as follows
\begin{equation}
T\left(\frac{\partial}{\partial T}\right)_{\mu,V} = -z \ln z \left(\frac{\partial}{\partial z}\right)_{\mu,V}.   
\end{equation}
By asking that Eq.s(10) and (11) coincide, it follows that
\begin{equation}
k_B T \left(\frac{\partial}{\partial T} \ln\mathcal{Q} \right)_{\mu,V} = -\bar{n}\frac{\mu}{T} + k_{B}\frac{\bar{N}}{k} T \left(\frac{\partial k}{\partial T} \right)_{V}
\end{equation}
i.e.\
\begin{equation}
T \left(\frac{\partial}{\partial T} \ln\mathcal{Q} \right)_{\mu,V} + \bar{n}\frac{\mu}{k _B T} = T \left(\frac{\partial}{\partial T} \ln\mathcal{Q} \right)_{\mu,V} + \bar{n} \ln z = \frac{\bar{N}}{k} T \left(\frac{\partial k}{\partial T} \right)_{V}.
\end{equation}
By applying finally the operator described by Eq.(12) to Eq.(13) and recalling that $z=\frac{\bar n}{\bar n + k}$, we find
\begin{equation}
\begin{split}
-z \ln z \left(\frac{\partial \ln\mathcal{Q}}{\partial z}\right)_{\mu,V} + \bar{n} \ln z & = -z \ln z \left[-k \frac{\partial \ln(1-z)}{\partial z} - \ln(1-z)\left(\frac{\partial k}{\partial z}\right)_{\mu,V}\right] + \bar{n} \ln z = \\
& = -z \ln z \left[(\bar n + k) - \ln(1-z)\left(\frac{\partial k}{\partial z}\right)_{\mu,V}\right] + \bar{n} \ln z = \\
& = z \ln z \ln(1-z)\left(\frac{\partial k}{\partial z}\right)_{\mu,V} = \\
& = \frac{\bar{N}}{k} T \left(\frac{\partial k}{\partial T} \right)_{V} = - \ln(1-z) T \left(\frac{\partial k}{\partial T} \right)_{V}
\end{split}
\end{equation}
and
\begin{equation}
-z \ln z \left(\frac{\partial k}{\partial z}\right)_{\mu,V} = T\left(\frac{\partial k}{\partial T}\right)_{V}.
\end{equation}
Being quantum effects in a boson system with fixed volume $V$ (i.e.\ constant $\bar n$) a low temperature phenomenon with the fugacity of the system
$z =\frac{\bar{n}}{\bar{n}+k} \rightarrow 1$, i.e.\ $k \rightarrow 0$, one has $\left(\frac{\partial k}{\partial T} \right)_{V} > 0$. It follows from Eq.(16) that also $\left(\frac{\partial k}{\partial z} \right)_{\mu,V} > 0$. \\
In addition $k$ is a canonical variable as it has been shown in \cite{thermo-clan} and it cannot depend on a grand canonical variable like the chemical potential $\mu$. \\
Now, willing to extract $k$ variable from $\left(\frac{\partial k}{\partial z} \right)_{\mu,V}$ in view of the \emph{canonical nature of} $k$, and to suppress the $z \ln z$ term in Eq.(16), we are led from
\begin{equation}
\left(\frac{\partial k}{\partial z} \right)_{\mu,V}=\left(\frac{\partial k \frac{{(\ln z)}^\alpha}{{(\ln z)}^\alpha}}{\partial z} \right)_{\mu,V}
\end{equation}
with constant $\alpha$, to assume that $k\,{(\ln z)}^\alpha$ is a function depending on volume $V$ and chemical potential $\mu$ only. \\
Accordingly, from (17) one has
\begin{equation}
\left(\frac{\partial k}{\partial z} \right)_{\mu,V}=\left(\frac{\partial k \frac{{(\ln z)}^\alpha}{{(\ln z)}^\alpha}}{\partial z} \right)_{\mu,V} =
k\,{(\ln z)}^\alpha \left(\frac{\partial \frac{1}{{(\ln z)}^\alpha}}{\partial z} \right) = - \frac{\alpha k}{z \ln z} > 0
\end{equation}
with $\alpha > 0$. Finally from (16) we find that $T$ dependence of $k$ is given by
\begin{equation}
T \left(\frac{\partial k}{\partial T} \right)_{V}=\alpha k.  
\end{equation}

Next step is to determine $V$ dependence of $k$ at fixed $T$. \\
By considering now the Helmoltz free energy $F$
$$
F = \mu \bar{n}-PV
$$
i.e.\ in case of the NB
\begin{equation}
F = k_{B}T \, \bar{n} \ln z - k_{B}T\bar{N},
\end{equation}
it follows
$$
\left(\frac{\partial F}{\partial V} \right)_{\bar{n},T} = -P
$$
which, being in this case $P\,V = k_{B}T\bar{N}$, implies
\begin{equation}
\left(\frac{\partial F}{\partial V} \right)_{\bar{n},T} = -\frac{1}{V}k_{B}T\bar{N} = \frac{1}{V}k_{B}T k \ln(1-z).
\end{equation}
Since now from Eq.(20) one has
\begin{equation*}
\begin{split}
\left(\frac{\partial F}{\partial V} \right)_{\bar{n},T} & = k_{B}T \, \bar{n} \left(\frac{\partial \ln z}{\partial V} \right)_{\bar{n},T} -
k_{B}T \left(\frac{\partial \bar N}{\partial V} \right)_{\bar{n},T} = \\
& = k_{B}T \,\bar{n} \, \frac{1}{z} \left(\frac{\partial z}{\partial V} \right)_{\bar{n},T} -
k_{B}T \left(\frac{\partial (-k \ln(1-z))}{\partial V} \right)_{\bar{n},T} = \\
& = k_{B}T \, \bar{n} \, \frac{1}{z} \left(\frac{\partial \frac{\bar n}{\bar n + k}}{\partial V} \right)_{\bar{n},T} +
k_{B} T \ln(1-z) \left(\frac{\partial k}{\partial V} \right)_{T} + k_{B} T \, k \left(\frac{\partial \ln(1-z)}{\partial V} \right)_{\bar{n},T} = \\
& = k_{B}T \, \bar{n}^2 \, \frac{1}{z} \left(\frac{\partial \frac{1}{\bar n + k}}{\partial V} \right)_{\bar{n},T} +
k_{B} T \ln(1-z) \left(\frac{\partial k}{\partial V} \right)_{T} - k_{B} T \, \frac{k}{1-z}\left(\frac{\partial z}{\partial V} \right)_{\bar{n},T} = \\
& = - k_{B}T \, \bar{n}^2 \, \frac{1}{z} {\left(\frac{1}{\bar n + k}\right)}^2  \left(\frac{\partial k}{\partial V} \right)_{T} +
k_{B} T \ln(1-z) \left(\frac{\partial k}{\partial V} \right)_{T} - \\
& \quad - k_{B} T \, (\bar n + k)\left(\frac{\partial z}{\partial V} \right)_{\bar{n},T},
\end{split}
\end{equation*}
it follows that 
\begin{equation}
\begin{split}
\left(\frac{\partial F}{\partial V} \right)_{\bar{n},T} & = - k_{B}T \, z \left(\frac{\partial k}{\partial V} \right)_{T} + k_{B} T \ln(1-z) \left(\frac{\partial k}{\partial V} \right)_{T} - k_{B} T \, (\bar n + k)\left(\frac{\partial z}{\partial V} \right)_{\bar{n},T} = \\
& = - k_{B}T \, z \left(\frac{\partial k}{\partial V} \right)_{T} + k_{B} T \ln(1-z) \left(\frac{\partial k}{\partial V} \right)_{T} - k_{B} T \, (\bar n + k)\left(\frac{\partial \frac{\bar n}{\bar n + k}}{\partial V} \right)_{\bar{n},T} = \\
& = - k_{B}T \, z \left(\frac{\partial k}{\partial V} \right)_{T} + k_{B} T \ln(1-z) \left(\frac{\partial k}{\partial V} \right)_{T} + k_{B}T \, z \left(\frac{\partial k}{\partial V} \right)_{T} = \\
& = k_{B} T \ln(1-z) \left(\frac{\partial k}{\partial V} \right)_{T}.
\end{split}
\end{equation}
The request that Eq.s(21) and (22) coincide, leads to
\begin{equation}
V \left(\frac{\partial k}{\partial V} \right)_{T} = k   
\end{equation}
which together with Eq.(19) gives
\begin{equation}
k = c\, T^\alpha\,V  
\end{equation}
with $c$ a constant independent on temperature $T$ and volume $V$, but depending on $\alpha$ only, i.e.\ a parameter which will be identified -- as will be shown in the following -- with the ideal clan gas degrees of freedom. It should be remarked that above results include the ultra-relativistic case as well of course the non-relativistic one, whereas the relativistic case is excluded in the present analytical approach and postponed to future work. \\
Notice that Eq.(24) is compatible with large $T$ values and $\bar N \rightarrow 1$ (i.e.\ an harder collision process with $k \rightarrow 0$) when the volume is becoming smaller ($V \rightarrow 0$) in a faster way than $T^{-\alpha}$ as the c.m.\ energy $\sqrt{s}$ becomes larger. \\
Equation (24) allows of course to determine all thermodynamical potentials given in \cite{thermo-clan} in terms of $c$, $T$ and $V$. It follows: \\
\emph{Entropy} $S$
\begin{equation}
\begin{split}
S & = k_{B}\bar{N}-\bar{n}\frac{\mu}{T}+k_{B}\frac{\bar{N}}{k} T \left(\frac{\partial k}{\partial T} \right)_{V} = \\
& = -k_{B} \left[(1+\alpha) \,c\,T^\alpha V \,\ln\left(\frac{c\,T^\alpha V}{\bar{n}+c\,T^\alpha V}\right)+\bar{n}\,\ln\left(\frac{\bar{n}}
{\bar{n}+c\,T^\alpha V}\right) \right];  
\end{split}   
\end{equation}
\emph{Helmoltz free energy} $F$
\begin{equation}
F = k_{B}T \left[\bar{n}\,\ln\left(\frac{\bar{n}}{\bar{n}+c\,T^\alpha V} \right)+c\,T^\alpha V \,\ln\left(\frac{c\,T^\alpha V}   
{\bar{n}+c\,T^\alpha V} \right) \right];     
\end{equation} 
\emph{Internal energy} $U$
\begin{equation}
U = -k_{B}T^2\,\ln(1-z)\left(\frac{\partial k}{\partial T} \right)_{V}= -k_{B} \,T\, \alpha \,k\, \ln(1-z)=\alpha\, k_{B} T \bar{N};
\end{equation} 
\emph{Specific heat at constant volume} $C_V$
\begin{equation}
C_V = k_{B} \,\alpha (\alpha+1) \,c\,T^\alpha V \,\ln\left(\frac{\bar{n}+c\,T^\alpha V}    
{c\,T^\alpha V} \right)-k_{B} \frac{\bar{n}}{\bar{n}+c\,T^\alpha V} \,{\alpha}^2 \,c\,T^\alpha V.     
\end{equation}

By comparing now the internal energy of an ideal gas $U=f(\nu)\,k_{B}T\ln\mathcal{Q}$ ($f(\nu)$ is here a function of the degrees of freedom of the system under investigation) with the result of our calculation (27) and by recalling that $\bar N = \ln\mathcal{Q}$ it follows that $\alpha=f(\nu)$, i.e.\ $\alpha$ is a function of the clan degrees of freedom $\nu$. \\
Notice that clan aggregation process is determined by the increase of the clan degrees of freedom, i.e.\ as the collision becomes harder clan degrees of freedom should be larger and a phase transition from a class of events to another one can occur by varying clan degrees of freedom. \\

\section{A phase transition between the semihard and hard component at parton level implies a discontinuity in the average number of particles of the two components at hadron level}

Within the framework discussed in the previous Section we would like to show: \\
I.\ that the expected phase transition in the ideal clan gas is a first order phase transition and that it is revealed at hadron level through a sudden discontinuity in the average number of final charged particles (the counterpart of the remark given in Ref.\cite{hard} that $P_{n}^{[NB]}$ vs.\ $n$ plot for the semihard class of events from concave $k_{semihard} >1$ is becoming convex with $k_{hard}<1$); \\
II.\ that in the maps of zeroes of the truncated NB (Pascal) MD grand canonical partition function in the rescaled complex $u$-plane $u=1$ is an accumulation point of zeroes (from inside for $k_{semihard}>1$ and from outside for $k_{hard}<1$ \cite{complex}). \\
Assume then that the \emph{partonic clan gas of the semihard component} in $pp$ collisions has below the critical c.m.\ energy, $\sqrt{s_{c}}$, $\nu_{semihard}$ degrees of freedom and above $\sqrt{s_{c}}$ the \emph{partonic clan gas of the hard component} has $\nu_{hard}$ degrees of freedom,
with of course $\alpha_{hard}=f(\nu_{hard}) > \alpha_{semihard}=f(\nu_{semihard})$. Recalling now that the truncated grand canonical partition function is analytic, the phase transition from the semihard to the hard ideal clan gas is expected to occur at parton level ($p$) in the thermodynamic limit, corresponding to
$$
\lim_{\bar{n}^{(semihard)}_p \rightarrow \infty \atop V^{(semihard)}_p \rightarrow \infty}{\frac{\bar{n}^{(semihard)}_p}{V^{(semihard)}_p}}=\rho^{(semihard)}_{c}=
\lim_{\bar{n}^{(hard)}_p \rightarrow \infty \atop V^{(hard)}_p \rightarrow \infty}{\frac{\bar{n}^{(hard)}_p}{V^{(hard)}_p}}=\rho^{(hard)}_{c},
$$
where $\rho^{(semihard)}_{c}$ and $\rho^{(hard)}_{c}$ are critical parton densities of the two phases. \\
In addition being the Helmoltz free energy density for the two phases
$$
f^{(i)}_p \equiv \lim_{V^{(i)}_p \rightarrow \infty}\frac{{F^{(i)}_p}}{V^{(i)}_p}
$$
with $i=semihard$ or $hard$, at the critical temperature $T_c$ one has
$$
f^{(semihard)}_p=f^{(hard)}_p
$$
i.e.\
$$
c_{\alpha_{semihard}}T_{c}^{\alpha_{semihard}}=c_{\alpha_{hard}}T_{c}^{\alpha_{hard}},
$$
which leads to
\begin{equation}
T_{c}=\left(\frac{c_{\alpha_{semihard}}}{c_{\alpha_{hard}}}\right)^\frac{1}{\alpha_{hard}-\alpha_{semihard}}.  
\end{equation}
Since $c_{\alpha_{semihard}}$ and $c_{\alpha_{hard}}$ do not depend on temperature $T$ and volume $V$, they are expected to be the same at hadron ($h$) and parton ($p$) level and by recalling Eq.(24), it follows that
\begin{equation*}
\begin{split}
T_{c} & = \lim_{V^{(semihard)}_{h} \rightarrow \infty \atop V^{(hard)}_{h} \rightarrow \infty}\left({\frac{\frac{k}{T_{h}^{\alpha_{semihard}} V^{(semihard)}_{h}}}{\frac{k}{T_{h}^{\alpha_{hard}} V^{(hard)}_{h}}}}\right)^\frac{1}{\alpha_{hard}-\alpha_{semihard}} = \\
& = T_{h}\lim_{V^{(semihard)}_{h} \rightarrow \infty \atop V^{(hard)}_{h} \rightarrow \infty}\left(\frac{V^{(hard)}_{h}}{V^{(semihard)}_{h}}\right)^\frac{1}{\alpha_{hard}-\alpha_{semihard}}.  
\end{split}   
\end{equation*}
A relation which by assuming that the particle densities are continuous also at hadron level,
$$
\lim_{\bar{n}^{(semihard)}_{h} \rightarrow \infty \atop V^{(semihard)}_{h} \rightarrow \infty}{\frac{\bar{n}^{(semihard)}_{h}}{V^{(semihard)}_{h}}}=\rho^{(semihard)}_{h}=
\lim_{\bar{n}^{(hard)}_{h} \rightarrow \infty \atop V^{(hard)}_{h} \rightarrow \infty}{\frac{\bar{n}^{(hard)}_{h}}{V^{(hard)}_{h}}}=\rho^{(hard)}_{h},
$$
implies 
\begin{equation}
T_{c} = T_{h} \lim_{\bar{n}^{(semihard)}_{h} \rightarrow \infty \atop \bar{n}^{(hard)}_{h} \rightarrow \infty} \left(\frac{\bar{n}^{(hard)}_{h}}{\bar{n}^{(semihard)}_{h}}\right)^\frac{1}{\alpha_{hard}-\alpha_{semihard}}.
\end{equation}
It follows from $T_c > T_{h}$ that Eq.(30) is certainly satisfied when
\begin{equation}
\lim_{\bar{n}^{(semihard)}_{h} \rightarrow \infty \atop \bar{n}^{(hard)}_{h} \rightarrow \infty}{\frac{\bar{n}^{(hard)}_{h}}{\bar{n}^{(semihard)}_{h}}>1}. 
\end{equation}
Therefore \emph{the phase transition} from the semihard ideal clan gas to the hard one, \emph{occurring at parton level, is revealed at hadron level} by a discontinuity in the average number of final particles of the two clan gas components. \\
By using GLPHD Eq.(31) can of course be rewritten as follows
\begin{equation}
\lim_{\bar{n}^{(semihard)}_{p} \rightarrow \infty \atop \bar{n}^{(hard)}_{p} \rightarrow \infty}{\frac{{\sigma}^{(hard)} \, \bar{n}^{(hard)}_{p}}{{\sigma}^{(semihard)} \, \bar{n}^{(semihard)}_{p}}>1}, 
\end{equation}
but since at the phase transition one has
$$
\bar{n}^{(semihard)}_{p} = \bar{n}^{(hard)}_{p}
$$
we get
\begin{equation}
\frac{{\sigma}^{(hard)}}{{\sigma}^{(semihard)}} > 1
\end{equation}
i.e.\
\begin{equation}
{\sigma}^{(hard)} > {\sigma}^{(semihard)}.
\end{equation}
Therefore Eq.(30) becomes
\begin{equation}
T_{c} = T_{h} \left(\frac{{\sigma}^{(hard)}}{{\sigma}^{(semihard)}}\right)^\frac{1}{\alpha_{hard}-\alpha_{semihard}}
\end{equation}
suggesting that hadronization process differs in the two classes of events, under the assumption that the hadronization temperature $T_h$ is the same. \\
In addition, being in the thermodynamic limit the entropy density $\frac{S}{V}$ equal to
\begin{equation}
\frac{S}{V} = -k_{B} \left[(1+\alpha) \,c\,T^\alpha \,\ln\left(\frac{c\,T^\alpha}{\rho_{c}+c\,T^\alpha}\right)+\rho_{c}\,\ln\left(\frac{\rho_{c}}{\rho_{c}+c\,T^\alpha}\right)\right]     
\end{equation}
it is clear that 
$$
\lim_{T \rightarrow T^{-}_c}\frac{S}{V} \neq \lim_{T \rightarrow T^{+}_c}\frac{S}{V}
$$
i.e.\ the phase transition from the semihard to the hard partonic ideal clan gas corresponds (being $S = -\left(\frac{\partial F}{\partial T} \right)_{\bar{n},V}$) to a first order phase transition.

Notice that the same conclusion can be achieved by studying the zero distribution properties of the $M$-truncated grand canonical partition function
\begin{equation}
\mathcal{Q}_M=1+z Z_{1}+z^2 Z_{2}+...+z^M Z_{M}.    
\end{equation}
From a known algebraic relation between coefficients and zeroes of a polynomial (see Ref.\cite{determinante}), in the thermodynamic limit ($M,V \rightarrow \infty$ with $M/V=\rho$) at the critical temperature $T_{c}$ one gets
\begin{eqnarray}
\sum_{n=1}^{M}z_{n} & = & -\frac{Z_{M-1}}{Z_{M}}=-\frac{M}{M+k-1}=-\frac{M}{M+c\,T^\alpha V-1} \stackrel{M,V \rightarrow \infty}{\longrightarrow} -\frac{\rho_{c}}{\rho_{c}+c\,T_{c}^\alpha} \nonumber \\
\sum_{n=1}^{M}{z}^2_{n} & = & \left(-\frac{Z_{M-1}}{Z_{M}}\right)^2-2\frac{Z_{M-2}}{Z_{M}} = \nonumber \\
& = & -\frac{{\left(\frac{M}{V}\right)}^2\left[\frac{2}{M V}+\left(\frac{M}{V}+c\,T^\alpha\right)\left(1-\frac{2}{M}\right)\right]}
{\left(\frac{M}{V}+c\,T^\alpha-\frac{2}{V}\right)
{\left(\frac{M}{V}+c\,T^\alpha-\frac{1}{V}\right)}^2}
\stackrel{M,V \rightarrow \infty}{\longrightarrow} -\left(\frac{\rho_{c}}{\rho_{c}+c\,T_{c}^\alpha}\right)^2 \nonumber \\
& \vdots & \nonumber \\
\sum_{n=1}^{M}{z}^M_{n} & \stackrel{M,V \rightarrow \infty}{\longrightarrow} & -\left(\frac{\rho_{c}}{\rho_{c}+c\,T_{c}^\alpha}\right)^M
\end{eqnarray} 
In Appendix it is shown that all the zeroes of a degree $M$ polynomial are distinct and lie on the circle of $1/{\gamma}$ radius at angles $\theta_{n}=n\frac{2\pi}{M+1}$ when
$$
\sum_{n=1}^{M}{z}^j_{n}=-\left(\frac{1}{\gamma}\right)^j, \;\; j=1,2,...,M.
$$
Notice that in the present approach, $1/{\gamma}$ is given by 
\begin{equation}
\frac{1}{\gamma}=\frac{\rho_{c}}{\rho_{c}+c\,T_{c}^\alpha},        
\end{equation} 
which corresponds to the critical fugacity $z_{c}$ and to a zeroes accumulation point, i.e.\ to a non-analicity point of the grand canonical partition function (4). Now, since the zeroes approach the positive real axis with a $\pi/2$ angle, it is confirmed that the phase transition is of first order \cite{Blythe-Evans}.
In fact it can be shown that the linear density of zeroes on the positive real axis is different from zero and equal to
$$
\lim_{M \rightarrow \infty} \frac{1/M}{z_{c}\,\theta_{1}}=\lim_{M \rightarrow \infty} \frac{1}{M}\frac{M+1}{2 \pi z_{c}}=\frac{1}{2 \pi z_{c}}. 
$$
Eq.(29) is obtained by using the identity
$$
\frac{\rho_{c}}{\rho_{c}+c_{\alpha_{semihard}}T_{c}^{\alpha_{semihard}}}=\frac{\rho_{c}}{\rho_{c}+c_{\alpha_{hard}}T_{c}^{\alpha_{hard}}},         
$$
i.e.\ by asking that $1/\gamma$ (the zeroes accumulation point) is the same for the two phases. \\
Notice that the properties of zeroes of the NB grand canonical partition function in the thermodynamic limit, Eq.s(38), do not depend on the shape of temperature dependence of parameter $k$. Therefore, the linear dependence of parameter $k$ on volume $V$, as described by Eq.(23), contains two information: 1) the presence of a phase transition (the zeroes stay on the circle of radius equal to the critical fugacity $z_c$, which corresponds to a zeroes accumulation point) and 2) the phase transition is of first order (since the zeroes stay on a circle, they approach the positive real axis with a $\pi/2$ angle \cite{Blythe-Evans}). \\
Notice that the general result given by Eq.(39) in the limit $\rho_c \rightarrow \infty$ implies $z_c \rightarrow 1$, as shown in Ref.\cite {complex}.

\section*{Appendix: On a theorem on the zeroes of a degree $M$ polynomial}

Let us consider a generic polynomial of degree $M$:
\begin{equation}
G_{M}(z)=\sum_{m=0}^{M}z^m P_{m}=P_{0}+z P_{1}+z^2 P_{2}+...+z^M P_{M}.  
\end{equation}
We now prove the following theorem: if the zeroes of the algebraic equation
\begin{equation}
G_{M}(z)=0           
\end{equation}
satisfy the conditions
\begin{equation}
\sum_{n=1}^{M}{z}^j_{n}=-\left(\frac{1}{\gamma}\right)^j, \;\; j=1,2,...,M
\end{equation}
they are all distinct and lie on the circle of $1/{\gamma}$ radius at angles $\theta_{n}=n\frac{2\pi}{M+1}$, with $n=1,2,...,M$. \\ 

Proof: since the zeroes of a polynomial occur in complex conjugate pairs, Eq.s(42) can be rewritten as follows
\begin{eqnarray}                                                                     
\sum_{n=1}^{M}z_{n}^j & = & \sum_{n=1}^{M} {\mid z_{n} \mid}^j (\cos[j \theta_{n}]+i \sin[j\theta_{n}])= \nonumber \\
& = &\sum_{n=1}^{M} {\mid z_{n} \mid}^j \cos[j\theta_{n}]=-\left(\frac{1}{\gamma}\right)^j, \;\; j=1,2,...,M.     
\end{eqnarray} 
The request that the zeroes of Eq.(41) lie on the circle of radius $1/{\gamma}$, implies of course that $\mid z_{n} \mid=1/\gamma, \forall n=1,2,...,M$,
and therefore the following condition for the $\theta_{n}$ angles holds:
\begin{equation}
\sum_{n=1}^{M} \cos[j\theta_{n}]=-1, \;\; j=1,2,...,M.   
\end{equation}

The theorem can be proved simply by showing that Eq.s(44) are indeed verified, as will be shown in the following. Notice first that
\begin{eqnarray*}
\sum_{n=0}^M \cos[nx] &=& \Re\left[\sum_{n=0}^M e^{inx}\right]=\Re\left[\frac{e^{i(M+1)x}-1}{e^{ix}-1}\right]= \\ \\ 
&=& \Re\left[\frac{e^{i(M+1)x/2}-e^{-i(M+1)x/2}}{e^{ix/2}-e^{-ix/2}} \frac{e^{i(M+1)x/2}}{e^{ix/2}}\right]= \\ 
&=& \frac{\sin[(M+1)x/2]}{\sin[x/2]}\,\Re[e^{iMx/2}] = \frac{\cos[Mx/2]\sin[(M+1)x/2]}{\sin[x/2]}
\end{eqnarray*}
and of course
$$
\sum_{n=1}^M \cos[nx]=\frac{\cos[Mx/2]\sin[(M+1)x/2]}{\sin[x/2]}-1.
$$
Calling now $nx=\theta_{n}$, with $\theta_{n}=n\frac{2\pi}{M+1}$, one gets
\begin{eqnarray*}
\sum_{n=1}^M \cos[\theta_{n}] & = & \sum_{n=1}^M \cos\left[n\frac{2\pi}{M+1}\right]=\frac{\cos[\frac{1}{2}M\frac{2\pi}{M+1}]\sin[\frac{1}{2}(M+1)\frac{2\pi}{M+1}]}{\sin[\frac{1}{2}\frac{2\pi}{M+1}]}-1= \\ 
& = &\frac{\cos[M\frac{\pi}{M+1}]\sin[\pi]}{\sin[\frac{\pi}{M+1}]}-1=-1.
\end{eqnarray*}
Accordingly, by writing $nx$ as $j\theta_{n}$, with $\theta_{n}=n\frac{2\pi}{M+1}$ and $j=1,2,...,M$, it follows that
\begin{eqnarray*}
\sum_{n=1}^M \cos[j\theta_{n}] & = & \sum_{n=1}^M \cos\left[jn\frac{2\pi}{M+1}\right]=\frac{\cos[\frac{j}{2}M\frac{2\pi}{M+1}]\sin[\frac{j}{2}(M+1)\frac{2\pi}{M+1}]}{\sin[\frac{j}{2}\frac{2\pi}{M+1}]}-1= \\ 
&=&\frac{\cos[jM\frac{\pi}{M+1}]\sin[j\pi]}{\sin[j\frac{\pi}{M+1}]}-1=-1,
\end{eqnarray*}
and therefore
$$
\sum_{n=1}^M \cos\left[jn\frac{2\pi}{M+1}\right]=-1,
$$
with $j=1,2,...,M$. 
Since the set of zeroes of a polynomial is univocally defined, the theorem is proved.

\end{document}